\documentclass[10pt,aps,prl,twocolumn,showpacs,amssymb,floatfix]{revtex4-1}
\usepackage{amsmath,graphics,epsfig,mathrsfs}
\usepackage{bm}
\usepackage{epigraph}

\newcommand{\nn}{\nonumber}

\usepackage{color}

\newcommand{\etal}{{\em et al.~}}

\begin{document}
\title{Magnon-Mediated Dzyaloshinskii-Moriya Torque in Homogeneous Ferromagnets}
\author{Aur{\' e}lien Manchon$^{1}$}
\email{aurelien.manchon@kaust.edu.sa}
\author{P. Birame Ndiaye$^{1}$}
\author{Jung-Hwan Moon$^2$}
\author{Hyun-Woo Lee$^{3}$}
\author{Kyung-Jin Lee$^{2,4}$}
\email{kj\_lee@korea.ac.kr}
\affiliation{$^1$ King Abdullah University of Science and Technology (KAUST), Physical Science and Engineering Division, Thuwal 23955-6900, Saudi Arabia.\\
$^2$Department of Materials Science and Engineering, Korea University, Seoul 136-701, Korea. \\
$^3$PCTP and Department of Physics, Pohang University of Science and Technology, Kyungbuk 790-784, Korea.\\
$^4$KU-KIST Graduate School of Converging Science and Technology, Korea University, Seoul 136-713, Korea.}\date{\today}

\begin{abstract}
In thin magnetic layers with structural inversion asymmetry and spin-orbit coupling, the Dzyaloshinskii-Moriya interaction arises at the interface. When a spin wave current ${\bf j}_m$ flows in a system with a homogeneous magnetization {\bf m}, this interaction produces an effective field-like torque of the form ${\bf T}_{\rm FL}\propto{\bf m}\times({\bf z}\times{\bf j}_m)$ as well as a damping-like torque, ${\bf T}_{\rm DL}\propto{\bf m}\times[({\bf z}\times{\bf j}_m)\times{\bf m}]$, the latter only in the presence of spin-wave relaxation (${\bf z}$ is normal to the interface). These torques mediated by the magnon flow can reorient the time-averaged magnetization direction and display a number of similarities with the torques arising from the electron flow in a magnetic two dimensional electron gas with Rashba spin-orbit coupling. This magnon-mediated spin-orbit torque can be efficient in the case of magnons driven by a thermal gradient.\end{abstract}
 
\maketitle
Recent developments in condensed matter physics have renewed the interest of the scientific community in the design and properties of materials with large spin-orbit coupling. Topics such as spin Hall effect \cite{she}, topological insulators \cite{ti}, or skyrmions \cite{sk}, all taking advantage of relativistic effects in solid state, have profoundly challenged our understanding of spin transport lately and present tremendously rich opportunities for innovative expansion of the research in condensed matter systems. Utilizing spin-orbit coupling to enable the electrical manipulation of ferromagnets and magnetic textures has attracted a considerable amount of interest in the past few years \cite{miron,liu,dms}. The key mechanism, tagged {\em spin-orbit torque}, appears in ultrathin magnetic systems displaying inversion symmetry breaking such as (but not limited to) bilayers composed of noble metals and ferromagnets. The recent experimental results are interpreted in terms of Rashba \cite{rashba} and spin Hall effect-induced torques \cite{she} and the complexity of the spin transport in such systems is currently under intense investigations \cite{manchon,rashba2,Stiles}. A major progress in this field has been to recognize the importance of Dzyaloshinskii-Moriya interaction (DMI) \cite{dm}. DMI results from the spin-orbit coupling in systems with broken inversion symmetry and is a necessary ingredient for the emergence of skyrmions and chiral spin textures \cite{skyrmions,bogdanov,chen}. Interestingly, DMI also arises from the interfacial spin-orbit coupling in ultrathin magnetic bilayers \cite{thiaville,kim} and results in chiral magnetic domain walls \cite{chen}, providing an explanation to mysterious experimental behaviors such as current-induced domain wall motion against the electron flow \cite{thiaville,parkin,emori}.

In conjunction with electrically driven {\em spin-orbit torques}, another adjacent emerging topic aims at exploiting magnon flows and propagating spin waves instead of electrical carriers \cite{magnonics}. Indeed, magnons can carry spin currents \cite{saitoh1}, transmit information \cite{saitoh2} and even control the motion of magnetic domain walls \cite{swdw,swdwth} and skyrmions \cite{prlsk}. The magnon flow may be driven by radio-frequency (RF) magnetic fields or temperature gradients \cite{hatami}, the latter being an important topic of the  {\em spin caloritronics} field \cite{spincaloritronics}. Recently, it has been realized that DMI impacts the propagation of spin waves just like spin-orbit coupling affects the electron flow, resulting in topological behaviors such as the magnon Hall effect and edge currents \cite{onose}. It was reported that the DMI effect on the spin wave dispersion is similar to the Rashba spin-orbit coupling effect on electron dispersion \cite{costa,Moon,swdft}. Therefore, one anticipates that the spin-orbit torque due to electron flow in Rashba spin-orbit coupled systems might have its counterpart due to magnon flow in systems displaying DMI.\par

\begin{figure}[ttbp]
\begin{center}
\psfig{file=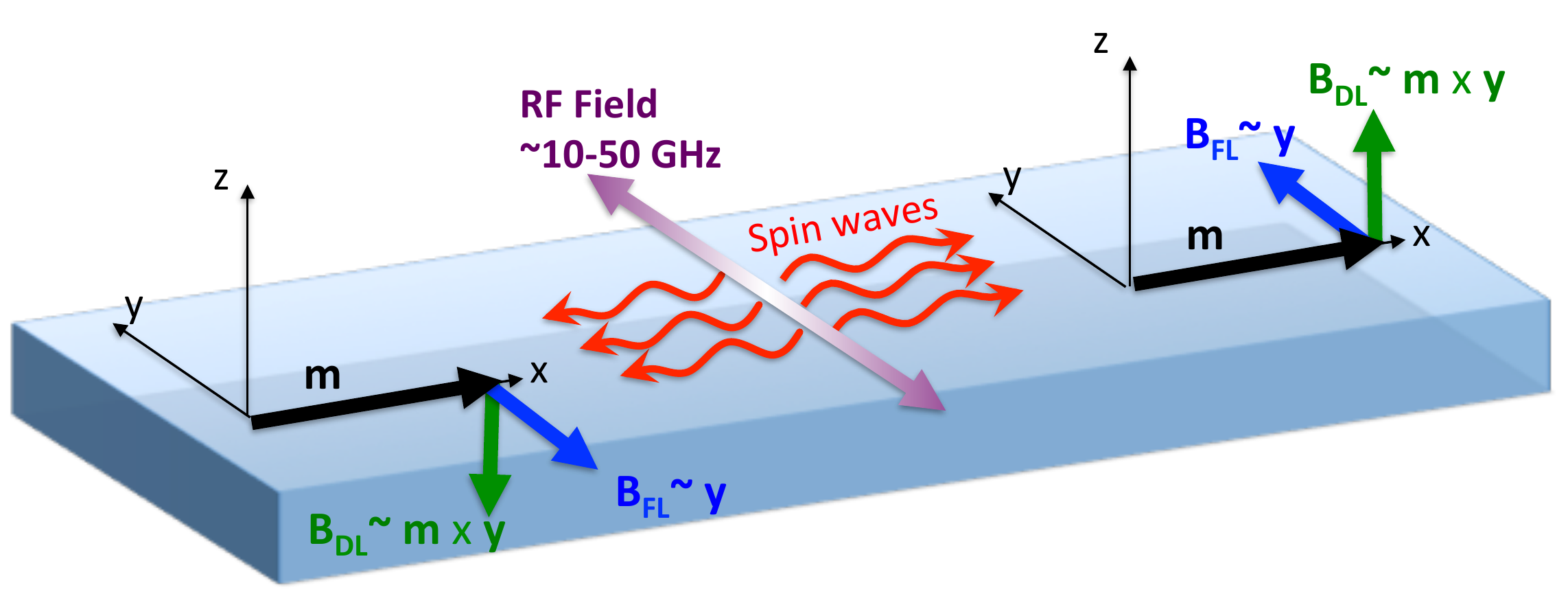,width=0.9\columnwidth} \caption{\label{fg1}
(Color online) Schematics of the in-plane magnetized stripe studied in this work. The magnetization is initially oriented along ${\bf x}$ and propagating spin waves are generated by an ac field applied in the center of the stripe at $x=0$. Due to Dzyaloshinskii-Moriya interaction, the spin wave flow induces effective fields, ${\bf B}_{\rm FL}\propto {\bf y}$ and ${\bf B}_{\rm DL}\propto {\bf m}\times{\bf y}$ resulting in small deviations of the background magnetization $\Delta m_{y,z}$.}
\end{center}
\end{figure}
In this Letter, we demonstrate that even in the absence of a magnetic texture, a magnon flow generates torques if magnons are subject to DMI just as an electron flow generates torques when submitted to Rashba interaction,  even when the magnetization is homogeneous \cite{manchon,rashba2}. A direct consequence is the capability to control the magnetization direction of a homogeneous ferromagnet by applying a temperature gradient or a local RF field to generate the magnon flow. We show that merging the spin-orbit torques with spin caloritronics is rendered possible by the emergence of DMI in magnetic materials and opens promising avenues in the development of chargeless information technology.\par

The magnon-induced torque arises both in longitudinal in-plane and perpendicular magnetic anisotropy systems. For simplicity, we demonstrate it only for the longitudinal in-plane magnetic anisotropy case. The other case is treated in Ref. \onlinecite{supp}. Let us consider a thin magnetic film with a magnetization aligned along the in-plane easy axis ($x$-axis) and subjected to an external ac magnetic field applied locally to make spin waves propagate along the $x$-axis, as displayed in Fig. \ref{fg1}. In this system, the magnetic energy reads 
\begin{eqnarray}\label{eq:w}
W&=&A\sum_i(\partial_i{\bf m})^2-D{\bf m}\cdot[({\bf z}\times{\bm \nabla})\times{\bf m}]\\
&&+2\pi M_s^2({\bf m}\cdot{\bf z})^2-K({\bf m}\cdot{\bf x})^2\nn,
\end{eqnarray}
where the first two terms are the symmetric exchange ($A$) and antisymmetric Dzyaloshinskii-Moriya ($D$) exchange energies, the last two terms are the demagnetizing ($2\pi M_s^2$) and the in-plane anisotropy ($K$) energies, and ${\bm\nabla}=(\partial_x,\partial_y,\partial_z)$. The form of the DMI we adopt here is derived for a cylindrically symmetric system with an interfacial inversion asymmetry along the normal ${\bf z}$ \cite{bogdanov,thiaville}. All along the present study, we consider that the DMI is smaller than a certain threshold value $D_c$ so that the uniformly magnetized state is energetically stable \cite{supp,Moon}. \par

Before going forward with the simulations and in order to get a quantitative understanding of the physics at stake, we first analytically derive the magnon-mediated Dzyaloshinskii-Moriya torque. We assume that in the absence of magnon flow, the magnetization is initially aligned along the in-plane anisotropy axis $\eta{\bf x}$ ($\eta=\pm1$). The magnon flow then induces a (dimensionless) deviation of {\bf m} from $\eta{\bf x}$. To describe this deviation, we express {\bf m} in spherical coordinates, ${\bf m}=s_r{\bf e}_r+s_\theta{\bf e}_\theta+s_\phi{\bf e}_\phi$, where ${\bf e}_r$ is a unit vector along the {\em time-averaged} magnetization direction and ${\bf e}_{\theta,\phi}\cdot{\bf e}_r=0$. Due to the spin wave, parametrized by ($s_\theta,s_\phi$), $s_r$ is smaller than 1 since $s_r^2 + s_\theta^2 + s_\phi^2 =1$. We show below that in the presence of the DMI, ${\bf e}_r$ deviates from $\eta{\bf x}$. We denote the deviation along {\bf y} and {\bf z} directions by $\Delta m_y$ and $\Delta m_z$, respectively. These deviations imply that the magnon flow generates effective magnetic fields along {\bf y} and {\bf z} directions, which in turn induce torques. Since their magnitudes are proportional to $D$, we call them DM torques. To determine $\Delta m_{y,z}$ \cite{supp}, one injects the spherical expression of {\bf m} into the Landau-Lifshitz-Gilbert (LLG) equation,  defined as $\partial_t{\bf m}=\gamma{\bf m}\times\partial_{\bf m}W+\alpha{\bf m}\times\partial_t{\bf m}$, and after averaging over time we obtain the differential equations describing the spatial variation of the deviations \cite{supp}
\begin{eqnarray}\label{eq:dmy}
&&\partial_x^2\Delta m_y-\frac{1}{\lambda^2}\Delta m_y=\frac{D^*}{J}\langle s_\phi\partial_x s_\theta\rangle,\\
&&\partial_x^2\Delta m_z-\frac{1}{\lambda_{\rm d}^2}\Delta m_z=-\eta\frac{D^*}{J}\langle s_\phi\partial_x s_\phi\rangle,\label{eq:dmz}
\end{eqnarray}
where $\lambda$ (=$\sqrt{J/H_{\rm k}}$) and $\lambda_{\rm d}$ (=$\sqrt{J/(H_{\rm k}+H_{\rm d})}$) are the in-plane and out-of-plane characteristic lengths of the magnetic texture, with  $J=2A/M_s$, $H_{\rm k}=2K/M_s$, $H_{\rm d}=4\pi M_s$, and $D^*=2D/M_s$. The right hand sides of Eqs. (\ref{eq:dmy}) and (\ref{eq:dmz}) show that the deviations $\Delta m_{y,z}$ are driven by DMI, mediated by propagating spin waves. The right hand sides are at least in the {\em second order} in spin wave amplitude. Thus to evaluate $\Delta m_{y,z}$ up to the same order, it suffices to evaluate $s_\theta$ and $s_\phi$ only up to the {\em first order} by using the linearized LLG equation,
\begin{eqnarray}\label{sphi}
&&\partial_t s_\theta+\alpha \partial_t s_\phi=\gamma J\partial_x^2 s_\phi-\gamma H_{\rm k}s_\phi,\\
&&\partial_t s_\phi-\alpha \partial_t s_\theta=-\gamma J \partial_x^2 s_\theta+\gamma (H_{\rm k}+H_{\rm d})s_\theta.\label{sz}
\end{eqnarray}
For the situation depicted in Fig. \ref{fg1} where the spin wave is generated at $x=0$ by a localized ac magnetic field, Eqs. (\ref{sphi}) and (\ref{sz}) yield a spin wave, $\psi_m=s_\theta+is_\phi$, of the form
\begin{eqnarray}\label{sw}
&&\psi_m=e^{-|x|/2\Lambda}[s_\theta^0\cos(q|x|-\omega t)+is_\phi^0\sin(q|x|-\omega t)],\\
&&\Lambda=\gamma J q/\alpha \omega,\; \omega =\gamma\sqrt{(Jq^2+H_{\rm d}+H_{\rm k})(Jq^2+H_{\rm k})},
\end{eqnarray}
where $2\Lambda$ is the spin wave attenuation length. Now, by inserting Eq. (\ref{sw}) into Eqs. (\ref{eq:dmy}) and (\ref{eq:dmz}), and taking the time average over a spin wave precession period (i.e. $\langle s_\phi\partial_xs_\phi\rangle=-\frac{{\rm sign}(x)}{4\Lambda}(s_\phi^0)^2e^{-|x|/\Lambda}$ and $\langle s_\phi\partial_xs_\theta\rangle=-\frac{{\rm sign}(x)q}{2}s_\phi^0s_\theta^0e^{-|x|/\Lambda}$),  we can track the impact of this damped spin wave on the deviations $\Delta m_{y,z}$. Considering that the right hand sides of Eqs. (\ref{eq:dmy}) and (\ref{eq:dmz}) are odd functions of $x$, $\Delta m_{y,z}$ should vanish at $x$=0. Combined with the boundary condition, $\Delta m_{y,z} \vert_{x \to \infty}=0$, one finds that 
\begin{eqnarray}\label{Deltamy}
\Delta m_y &=&-{\rm sign}(x)\frac{H_{\rm DMF}^{\rm eff}}{H_{\rm k}}\frac{\Lambda^2}{\Lambda^2-\lambda^2} (1-e^{-|x|/\lambda^*}),\\
\Delta m_z &=&\eta{\rm sign}(x) \frac{H_{\rm DMD}^{\rm eff}}{H_{\rm k}+H_{\rm d}}\frac{\Lambda^2}{\Lambda^2-\lambda_{\rm d}^2} (1-e^{-|x|/\lambda_{\rm d}^*}),\label{Deltamz}
\end{eqnarray}
where $H_{\rm DMF}^{\rm eff} = D^* q s_\theta^0 s_\phi^0e^{-|x|/\Lambda}/2$ and $H_{\rm DMD}^{\rm eff} = D^* (s_\phi^0)^2e^{-|x|/\Lambda}/4\Lambda$ and $\lambda_{\rm{(d)}}^{*-1}=\lambda_{\rm{(d)}}^{-1}-\Lambda^{-1}$. Note that $\Delta m_{y,z}$ are proportional to $D$ and spin wave amplitude square, implying that the magnon flow generates effective magnetic fields along {\bf y} and {\bf z} directions. Considering that ${\bf m}\approx\eta{\bf x}$, the two field directions may be represented as {\bf y} and ${\bf y}\times{\bf m}$, consistently with the absence (presence) of the factor $\eta$ in Eq. (\ref{Deltamy}) [Eq. (\ref{Deltamz})]. These fields generate the {\em DM field-like torque} (FLT) $\propto {\bf m}\times{\bf y}$ and the {\em DM damping-like torque} (DLT) $\propto {\bf m}\times({\bf y}\times{\bf m})$, in complete analogy with the Rashba torque \cite{manchon,rashba2}. Note also that $\Delta m_{y,z}$ are proportional to sign$(x)$, implying that in the two regions, $x>$0 and $x<$0, where the spin wave propagates in the opposite directions, the effective field signs are opposite. Thus the vectors {\bf y} and ${\bf y}\times{\bf m}$ for the fields actually amount to ${\bf z} \times {\bf j}_m$ and  $({\bf z} \times {\bf j}_m)\times{\bf m}$, respectively, where ${\bf j}_m$ is the spin wave current.

To get a further insight into the impact of propagating spin waves on the otherwise spatially homogeneous background magnetization, we now show micromagnetic simulation results for a semi-one dimensional system (i.e., the system is discretized along the length direction with the unit cell size of 4 nm - total length of 16 $\mu$m -, but not along the width or the thickness direction). We solve the LLG equation with the magnetic energy function given in Eq. (\ref{eq:w}). We define the gyromagnetic ratio $\gamma$ = 1.76$\times$10$^7$ Oe$^{-1}$s$^{-1}$, the saturation magnetization $M_s$ = 800 emu/cm$^3$, the exchange stiffness constant $A$ = 1.3$\times$10$^{-6}$ erg/cm, and vary the easy axis anisotropy field $H_{\rm k}$, the demagnetization field along the thickness direction $H_{\rm d}$, and the damping constant $\alpha$. To excite spin waves, we apply an ac field $H_{\rm ac}$$\cos (2 \pi f t) \bf{ y}$ to two unit cells at the center of the model system ($x$ = 0) where $H_{\rm ac}$=100 Oe. This choice of the localized ac field is consistent with the situation assumed for Eqs. (\ref{Deltamy}) and (\ref{Deltamz}). We  consider the absorbing boundary condition~\cite{Berkov,Seo} at the system edges to suppress spin wave reflection.\par

\begin{figure}[ttbp]
\begin{center}
\psfig{file=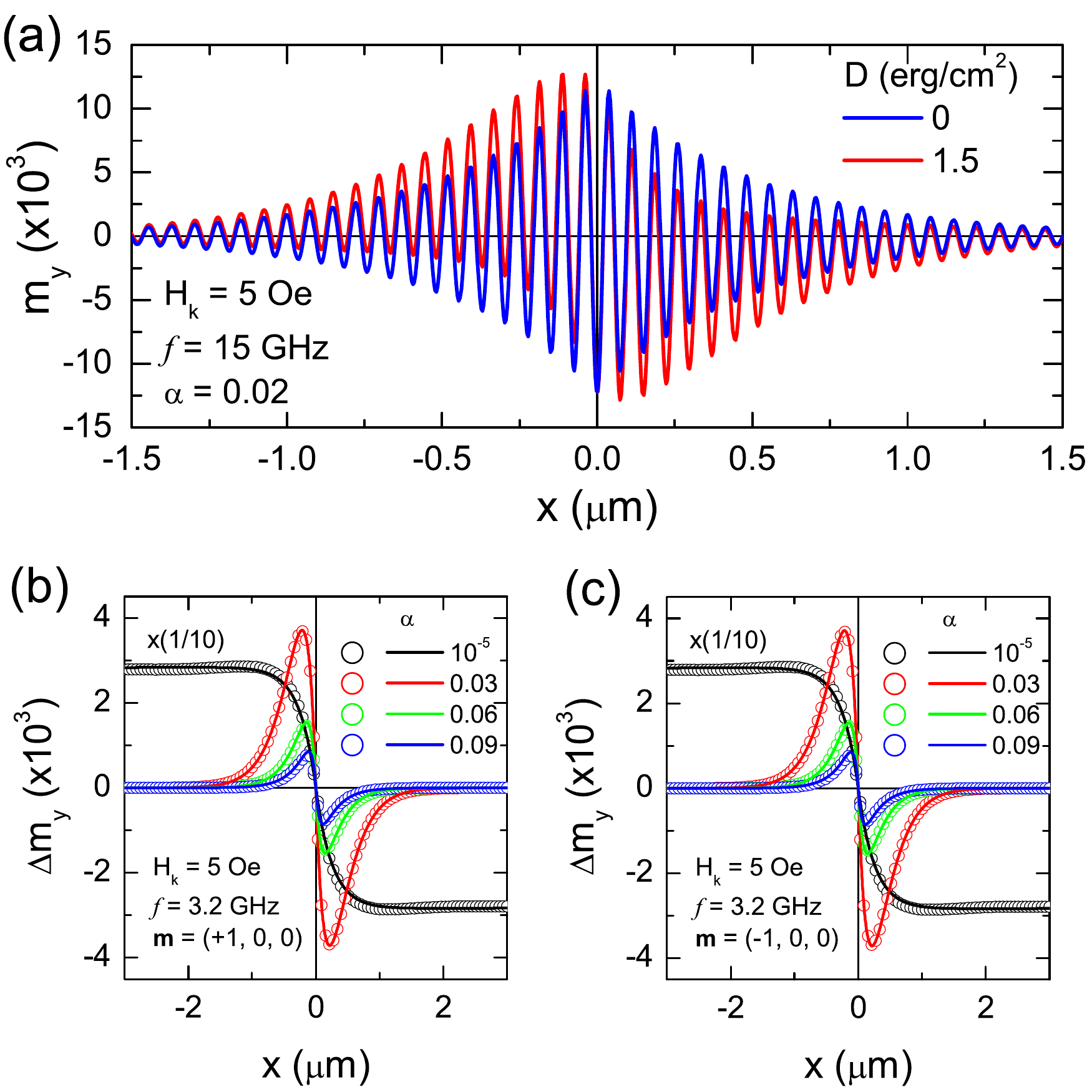,width=1.0\columnwidth} \caption{\label{fg2}
(Color online) Numerical results for magnon-mediated Dzyaloshinskii-Moriya field-like torque. (a) Spatial distribution of the normalized $y$-component of magnetization (= $m_y$) for $D$=0 (blue) and $D$=1.5 erg/cm$^2$ (red). (b, c) Normalized magnetization tilting $\Delta m_y$ for various damping constants $\alpha$ when the magnetization initially lies along $+\bf{x}$ (b) and $-\bf{x}$ (c), for $D$=1.5 erg/cm$^2$ calculated numerically (open symbols) and using Eq. (8) (solid lines). In (b, c), the results with $\alpha=10^{-5}$ are multiplied by 1/10.}
\end{center}
\end{figure}

 Figure~\ref{fg2}(a) shows the spatial distribution of the transverse projection of the magnetization direction $m_y$ for different DMI coefficients $D$. For $D=0$, the spatial distribution of $m_y$ is {\em symmetric} with respect to the spin wave source ($x$ = 0) and described by $m_y\cong 0+s_y(x) \cos(q|x|-\omega t)$, where $"0"$ represents the $y$-component of the background magnetization and the spin wave amplitude $s_y(x)$ decays with growing $|x|$ due to damping. For $D \neq 0$, on the other hand, the distribution acquires an {\em anti-symmetric} component $ \Delta m_y(x)$ with respect to $x$=0 and $m_y \cong \Delta m_y(x) + s_y(x) \cos (q|x|-\omega t)$, according to Fig. \ref{fg2}(a). Thus the propagating spin wave modifies the y-component of the background magnetization from 0 to a nonzero time-averaged value $\Delta m_y(x)$, which is shown in Fig. \ref{fg2}(b) for a background magnetization initially lying along $+{\bf x}$. Reversing the direction of the background magnetization along $-{\bf x}$ does not change the sign of $\Delta m_y$ [see Fig. \ref{fg2}(c)]. 
 Thus the simulation result for $\Delta m_y$ is qualitatively consistent with the vector expression - ${\bf z}\times{\bf j}_m$ obtained from Eq. (\ref{Deltamy}). For a quantitative comparison, Figs.~\ref{fg2}(b) and (c) show the deviation $\Delta m_y$ for various values of $\alpha$, where symbols are numerical results and lines are obtained from Eq. (\ref{Deltamy}). When $s_\theta^0$ and $s_\phi^0$ in $H_{\rm DMF}^{\rm eff}$ are determined from the numerical simulation result at $x$=0, the analytical expression reproduces the numerical results very well for entire range of $x$. The result for vanishingly small $\alpha$ (=$10^{-5}$) is worth emphasis. In this case, $\Delta m_y$ becomes flat and finite away from $x$=0. Thus the DM-FLT does not require any spatial variations of the background magnetization profile (e.g. domain walls), which differs from the spin-wave-induced torque in the absence of the DMI \cite{swdwth}. \par
 
\begin{figure}[ttbp]
\begin{center}
\psfig{file=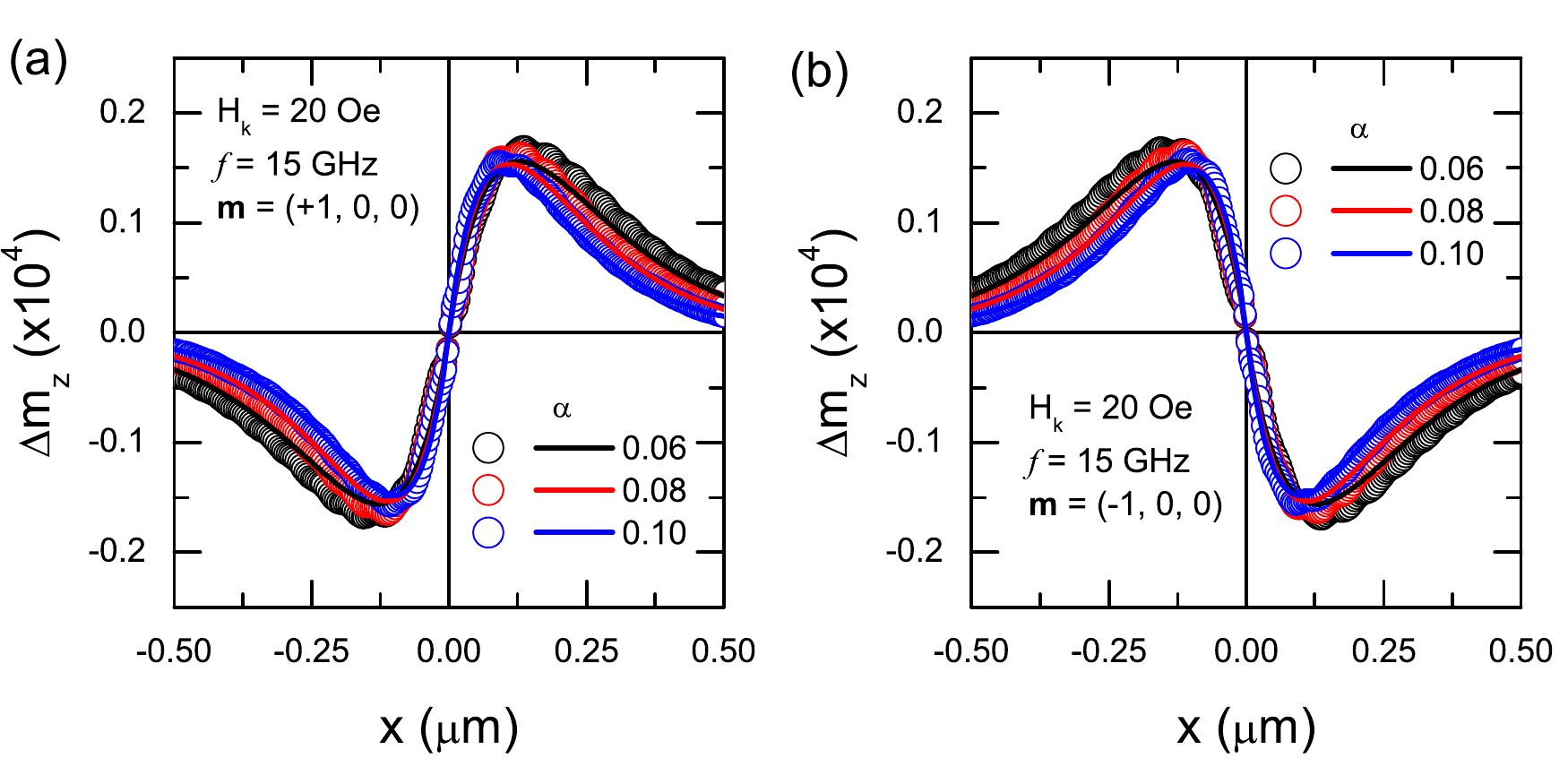,width=1.0\columnwidth} \caption{\label{fg3}
(Color online) Numerical results for magnon-mediated Dzyaloshinskii-Moriya damping-like torque. (a, b) Normalized magnetization tilting $\Delta m_z$ for various damping constants $\alpha$ when the magnetization initially lies along $+\bf{x}$ (a) and $-\bf{x}$ (b), for $D$=1.5 erg/cm$^2$ calculated numerically (open symbols) and using Eq. (9) (solid lines). Here we assume $H_d$=0.}
\end{center}
\end{figure}

Another intriguing observation is the emergence of the DM-DLT [$\propto{\bf m}\times({\bf y}\times {\bf m})$] that induces an out-of-plane deviation $\Delta m_z$. Figures~\ref{fg3}(a) and (b) show the spatial distribution of $\Delta m_z$ for different $\alpha$ when the magnetization is initially aligned along +{\bf x} and -{\bf x}, respectively. For clarity, we assume $H_{\rm d} = 0$ to make $\lambda_{\rm d}$ and $\Delta m_z$ larger. The numerical results (open symbols) are in good agreement with Eq. (\ref{Deltamz}) (solid lines) and $\Delta m_z$ consistently changes sign with the magnetization direction. As demonstrated by Fig.~\ref{fg3} and Eq. (\ref{Deltamz}), the DM-DLT is proportional to the damping constant $\alpha$ (since $\Lambda \propto 1/\alpha$), which echoes the non-adiabatic correction to the electronic spin torque in the presence of spin-flip relaxation, as proposed by Zhang \etal in magnetic textures and spin-valves \cite{zlf}. In metallic systems, the spin relaxation modifies the spin dynamics of the itinerant electrons which results in an additional torque component of the form $-\beta{\bf m}\times{\bm \tau}$, where ${\bm\tau}$ is the torque in the absence of spin relaxation and $\beta$ is proportional to the spin relaxation rate \cite{zlf}. The same effect is at the origin of the non-adiabatic torque in electron-driven and magnon-driven magnetic excitations: the magnetic damping $\alpha$ not only attenuates the spin wave current, but also relaxes the spin polarization carried by the spin waves producing the additional damping-like torque proportional to $H_{\rm DMD}^{\rm eff}$. As a result, the overall magnitude of $\Delta m_{z}$ vanishes in the limit of zero damping (see Ref. \onlinecite{supp}).


So far we have discussed the effect of the spin wave $\psi_m=s_\theta +i s_\phi$ on $\Delta m_{y,z}$. Still further insight can be gained by considering the effect of $\Delta m_{y,z}$ on $\psi_m$. For this, we go beyond the linearized LLG equation and introduce to Eqs. (\ref{sphi}) and (\ref{sz}) the lowest order coupling terms between $\Delta m_{y,z}$ and $\psi_m$, which are linear to both $\Delta m_{y,z}$ and $\psi_m$. In the short wave length regime, where the exchange and DM interactions dominate over the anisotropy, the resulting equations fall into the following form of the effective Schr\"odinger equation
\begin{equation}\label{eq:hm}
i\hbar\partial_t\psi_m=\hat{H}_m\psi_m=\left(\frac{\hat{\bf p}^2}{2m^*}+\frac{\alpha_{\rm DM}}{\hbar}\hat{\bf p}\cdot({\bf z}\times{\bf m})\right)\psi_m,
\end{equation}
where $\hat{\bf p}=-i\hbar{\bm \nabla}$ is the momentum operator, $m^*=\hbar M_s/4\gamma A$ is the magnon mass and $\alpha_{\rm DM}/\hbar=2\gamma D/M_s$ is Dzyaloshinskii-Moriya velocity for the spin waves. This equation instructively resembles Schr\"odinger's equation of an itinerant electron spin in a homogeneous magnetic two dimensional electron gas in the presence of Rashba spin-orbit coupling \cite{rashba}
\begin{equation}\label{eq:h}
i\hbar\partial_t\psi_e=\left(\frac{\hat{\bf p}^2}{2m}+\frac{\alpha_{\rm R}}{\hbar}\hat{\bf p}\cdot({\bf z}\times\hat{\bm \sigma})+J_{\rm ex}{\bf m}\cdot\hat{\bm \sigma}\right)\psi_e,
\end{equation}
where $\alpha_{\rm R}$ is the Rashba spin-orbit coupling and $J_{\rm ex}$ is the s-d exchange between itinerant electron spins $\hat{\bm\sigma}$ and the local moments aligned along ${\bf m}$. Equations (\ref{eq:hm}) and (\ref{eq:h}) differ by the presence of the s-d exchange term. Indeed, in contrast with electron spins, the magnon spin is by definition aligned along the local magnetization direction and its wavefunction $\psi_m$ is not a two-component spinor. Nevertheless, their similarity implies that properties of the Rashba system, such as current-induced Rashba field (also called {\em inverse spin galvanic effect}) of the form  ${\bf H}_{\rm R}=\alpha_{\rm R}m{\bf z}\times{\bf j}_s/\hbar M_s$, (${\bf j}_s$ being the flowing spin current) \cite{manchon}, are at least partly enabled by the presence of DMI in magnonic systems. The propagating spin wave and background magnetization ${\bf m}$ interact through the energy term $(\alpha_{\rm DM} / \hbar) \langle \hat{\bf p} \rangle_m\cdot({\bf z} \times{\bf m})$, where $\langle...\rangle_m$ denotes the quantum average on the magnon state $\psi_m$. This interaction term yields a torque of the form
\begin{eqnarray}\label{eq:torque}
{\bf T}_{\rm FL}=\gamma{\bf m}\times\partial_{\bf m}\langle\hat{H}_m\rangle=- {\bf m}\times  \{\frac{\alpha_{\rm DM}}{ \hbar} {\bf z} \times\langle{\bf p}\rangle_m \}
\end{eqnarray}
The expression within the brackets $\{...\}$ is nothing but the effective field of the DM-FLT, ${\bf H}_{\rm DMF}$, and $\langle{\bf p}\rangle_m$ amounts to the spin wave current. The DM-DLT can be obtained qualitatively by considering the correction due to the damping on the spin wave dynamics [see Eqs. (\ref{sphi})-(\ref{sz})]. In a Landau-Lifshitz approach, the magnetic damping corrects the torque by adding a contribution of the form $\gamma\alpha {\bf m}\times({\bf m}\times {\bf H}_{\rm DMF})$ that produces the DM-DLT term. \par


We next discuss the DM torque arising from the flow of magnons generated by the thermal gradient. Such a magnon flow is free from two drawbacks of the magnon flow generated by RF field: (i) the wavelength $2\pi/q$ of the spin wave is rather large ($74$ nm in the present study) producing a very small effective field [$H_{\rm DMF}^{\rm eff}<$0.15 Oe in Fig. (\ref{fg2})] and (ii) the magnon flow (hence, the DM torque) vanishes away from the RF source over the attenuation length $\Lambda$. Therefore, {\em thermal magnons} driven by a uniform temperature gradient, ${\bm\nabla} T$, are interesting candidates for the proposed effect. Following Eq. (\ref{eq:torque}), these magnons exert a torque of the form ${\bf T}_{\rm FL}^{\rm th}=(\alpha_{\rm DM}m^*/\hbar M_s){\bf m}\times({\bf z}\times{\bf j}_m)$ on the magnetization, where ${\bf j}_m$ is the magnon flow and can be estimated using a phenomenological Boltzmann equation \cite{supp,shulei} 
\begin{equation}
{\bf j}_m=\frac{k_B\partial_xT}{2\pi\alpha}\left[\int_{KV/k_BT}^{T_c/T}\frac{xe^xdx}{(e^x-1)^2}\right]{\bf x}.
\end{equation}
To evaluate the amount of magnonic current flowing through the system, we choose reasonable parameters for YIG as found in the literature \cite{swdw,Schreier}: $\alpha=10^{-4}$,  $V\approx1.9$nm$^3$, $K\approx 2\times 10^5$ erg/cm$^3$, $T_c=550$K, $T=300$K, and $\partial_x T=20$K/mm. Under these conditions, for a 1nm-thick magnetic slab, the flow of magnons is $j_m\approx2.29\times 10^{24}$s$^{-1}$.cm$^{-1}$, two orders of magnitude smaller than the usual critical switching current in conventional spin transfer torque configuration ($\approx 10^{26}$s$^{-1}$.cm$^{-2}$). The effective magnetic field generated by this magnon flow is $H_{\rm FL}=\hbar D/(2AM_s)j_x\approx1.7$ Oe, which is at least one order of magnitude larger than the effective field obtained from RF spin waves in this work. This conservative estimation leaves plenty of room for improvement such as (i) increasing the temperature gradient, (ii) reducing the magnetization damping, (iii) decreasing the magnetic anisotropy, and (iv) increasing the DMI. Tuning the parameters within a reasonable range \cite{supp}, it is possible to obtain $\Delta m_{y,z}$ up to 1\% which should be observable experimentally. Indeed, Fan et al. \cite{fan14}, measured a magnetization tilting of 0.1\% using MOKE, corresponding to a sensitivity of about 1$\mu$-rad while Xia et al. \cite{xia} showed a MOKE sensitivity of 60 $n$-rad.

In the present Letter, we demonstrated that Dzyaloshinskii-Moriya interaction mediated by spin waves can generate a torque on a homogeneous magnetization that resembles the Rashba torque, its electronic counterpart, displaying both field-like and damping-like components. The torque is expected to be much more efficient in the case of a magnon flow driven by a thermal gradient than for a standard RF-excited spin wave. It is important to stress out that our results are not limited to systems displaying {\em interfacial} DMI but can be also extended to materials accommodating {\em bulk} DMI since the energy functional needs only to display an antisymmetric exchange term $\sum_{ij}{\cal D}_{ij}\cdot{\bf S}_i\times{\bf S}_j$, such as in pyrochlore crystals \cite{onose} and chiral magnets \cite{tokura}. The present results build up a bridge between spin-orbit transport, magnonics and spin caloritronics and is expected to be detectable in systems ranging from thin magnetic bilayers to skyrmion crystals.\par

The authors acknowledge fruitful discussions with M.D. Stiles and K.W. Kim. K. J. L. was supported by NRF (NRF-2013R1A2A2A01013188), KU-KIST School Joint Research Program, and Human Resources Development Program, MKE/KETEP (No. 20114010100640). H. W. L. was supported by NRF (NRF-2013R1A2A2A05006237 and 2011-0030046).

\end{document}